\documentclass{article}

\usepackage{arxiv}

\usepackage[utf8]{inputenc} 
\usepackage[T1]{fontenc}    
\usepackage{hyperref}       
\usepackage{url}            
\usepackage{booktabs}       
\usepackage{amsfonts}       
\usepackage{nicefrac}       
\usepackage{microtype}      
\usepackage{lipsum}
\usepackage{graphicx}
\usepackage{caption}
\usepackage{subfigure}
\usepackage{float}
\usepackage{xspace}
\usepackage{underscore}
\usepackage{bm}
\DeclareUnicodeCharacter{2212}{\ensuremath{-}}
\graphicspath{ {./images/} }

\title{Prediction of Geometric Transformation on cardiac MRI via Convolutional Neural Network}

\author{
Xin Gao \\
  Department of Applied Mathematics\\
  DongHua University \\
  Shanghai, 201620, P.R. China \\
  \texttt{sanjin@mail.dhu.edu.cn} \\
 }
 
\begin{document}
\maketitle
\begin{abstract}

In the field of medical image, deep convolutional neural networks(ConvNets) have achieved great success in the classification, segmentation, and registration tasks thanks to their unparalleled capacity to learn image features. However, these tasks often require large amounts of manually annotated data and are labor-intensive. Therefore, it is of significant importance for us to study unsupervised semantic feature learning tasks. In our work, we propose to learn features in medical images by training ConvNets to recognize the geometric transformation applied to images and present a simple self-supervised task that can easily predict the geometric transformation. We precisely define a set of geometric transformations in mathematical terms and generalize this model to 3D, taking into account the distinction between spatial and time dimensions. We evaluated our self-supervised method on CMR images of different modalities (bSSFP, T2, LGE) and achieved accuracies of 96.4\%, 97.5\%, and 96.4\%, respectively. The code and models of our paper will be published on: 
\href{https://github.com/gaoxin492/Geometric_Transformation_CMR}{https://github.com/gaoxin492/Geometric\_Transformation\_CMR}

\end{abstract}


\section{Introduction}

In the field of medical image processing, the widespread adoption of deep convolutional neural networks\cite{lenet} has achieved great success in segmentation and registration tasks. For example, by training ConvNets for mammographymass lesion classification\cite{intro2}, biomedical image segmentation\cite{intro1}, and identification of the markers at the end of the femur\cite{intro3} with a massive amount of manually labeled data, they managed to learn the representation of different visual features in medical images. However, it is expensive for these supervised learning tasks to obtain ground truth, which requires intensive manual labeling effort. Therefore, we use data augmentation methods to expand the dataset before a supervised learning task, like image flipping, gamma correction, noise injection, PCA color augmentation, rotation, and scaling.

These operations are based on the premise that image enhancement only increases the number of samples and does not change the meaning of the image itself, also called geometric transformation invariance in \cite{CNN}. To better illustrate this premise, a type of self-learning supervised learning that defines an annotation-free pretext task is proposed to capture feature learning using only the visual information of the images or videos. Among them, the reorganization of geometric transformation that may be applied to the images is of particular interest for their simplicity and ease of manipulation without changing the morphology and content of the images. For example, Gidaris et al.\cite{Rotation} trained the RotNet( a specially designed ConvNet) to recognize the 2d rotation that is applied to the image that it gets as input, Lin et al.\cite{shou} trained ResNet to tell whether an image has been mirrored, Mumunl et al.\cite{CNN} had summarized some specialized CNN architectures for geometric transform-invariant representation. Medical images have unique characteristics compared to natural images, e.g., a data set contains only the same organs or parts and is dominated by grey-scale images.

When recognizing geometric transformations in medical images, it may only be necessary for the neural network to extract the contour and shape of the image, distinguish between light and dark, the relative position of organs, etc. However, in a more complex natural image dataset containing people, animals, objects, and landscapes, the neural network may need to extract other features such as the horizon, sky, shadows, relative positions of limbs, and facial features in addition. It is, therefore, a much easier task to identify geometric transformations in medical images than in natural images, and it is also easier to achieve a high level of accuracy. 

In addition to the artificial geometric transformations we make for the classification task in semi-supervised learning, the CMR images may also be stored in different orientations when actually shot. Therefore, in this paper, we train a ConvNet to recognize geometric transformations in CMR images not only to facilitate the correction of images to human vision-friendly orientations, as Zhang et al. \cite{zk} did, but also to explore the representation of medical image features by neural networks.

To be more specific, the main steps in our work are as follows. Firstly, we define a set of different geometric transformations. Each of them is applied to each image on the dataset, and the produced transformed images are fed to the ConvNet model that is trained to recognize the transformation of it. In this formulation, the set of geometric transformations defines the classification pretext task that the ConvNet model has to learn. We propose to limit the geometric transformations as the combined effect of mirror flips and rotations by multiples of 90 degrees. In the 2D case, the ConvNet model is trained on the 8-class image classification task (see Figure\ref{img1}). We have also extended this theory to the 3D case. We have suggested an intuition that an added third dimension would be more attractive to discuss if it were a time dimension than a spatial dimension. We trained this classification learning task on the MyoPS dataset in the experimental part. We obtained a high accuracy rate, showing that even though our approach is straightforward, it has still detected feature learning in neural networks.

Our contributions are:

\begin{enumerate}
    \item We propose more normatively a set of 2D geometric transformations containing mirror flips and rotations and extend it to 3D. And we come up with the idea of adding a time dimension rather than a spatial one.
    \item We propose a simple self-supervised task that can easily predict the geometric transformation on CMR images, which may inspire the design of tasks aimed at capturing other feature learning.
    \item We evaluated our self-supervised method on CMR images of different modalities and achieved high accuracy.

\end{enumerate}

\section{Methodology}
\subsection{Overview}
Our work aims to learn ConvNet-based semantic features of medical images employing unsupervised learning. This is based on the intuition that some neurons of a convolutional neural network can detect different features of an image in tasks such as recognition, segmentation, and registration. In particular, we focus on the feature of geometric transformations in images. Based on the study by Gidaris et al.\cite{Rotation}. On medical CMR images, we likewise train a ConvNet model $F(\cdot)$ to estimate more types of geometric transformation applied to a medical image given to it as an input. 

Consider a set of m different geometric transformations $G=\{g_y(\cdot)\}_{y=1}^m$ , which we will specify in the following subsection. In this set, $g_y(\cdot)$ is the operator that applies to image $X$ the geometric transformation with label $y$ that yields the transformed image $X^y = g_y(X) $. The ConvNet model $F(\cdot)$ gets as input an image $X^{y^*}$ (where the label $y^*$ is unknown and should be predicted by $F(\cdot)$). As with the usual classification task, the model outputs a probability distribution over all possible geometric transformations:
$$
F\left(X^{y *} \mid \theta\right)=\left\{F_y\left(X^{y *} \mid \theta\right)\right\}_{y=1}^m
$$
Where $F_y\left(X^{y *} \mid \theta\right)$ is the predicted probabilitiy for the geometric transformation with label y and $\theta$ are the learnable parameters of model $F(\cdot)$.

Therefore, given a set of $N$ training images $D=\left\{X_i\right\}_{i=0}^N$, the self-supervised training objective that the ConvNet model must learn to solve is:

$$
\min _{\theta} \frac{1}{N} \sum_{i=1}^N \bm{loss} (X_i, \theta),
$$

where the loss function $\bm{loss}(\cdot)$ is defined as:

$$
\bm{loss}\left(X_i, \theta\right)=-\log (\frac{\exp (y)} { \sum_{y=1}^m \exp ( (F_y({X_i}^{y^* }\mid \theta)))})
$$

\subsection{Geometric Transformation}

To avoid distorting the objects in the image, we propose to define the set of geometric transformations $G$ as the combined effect of mirror flip and rotations by multiples of 90 degrees. An additional advantage is that they do not leave visual artifacts on the transformed images that will lead the ConvNet to learn trivial features with no particle value for the visual perception tasks. 

For a point $p_i(x_i, y_i)$ in the 2D rectangular coordinate system, we denote the set of points $S = \{p_1,p_2,…,p_n\}$ by an $n\times2$ matrix $T_0= \left[\begin{array}{cc}x_1 & y_1 \\ x_2 & y_2 \\ \vdots & \vdots \\ x_n & y_n\end{array}\right]$. We can easily obtain the result of this point set after the mirror flip and rotation transformation by right multiplying the matrix $T_0$ by a $2\times2$ matrices in set $T=$ 

$$\{   \left[\begin{array}{ll}
1 & 0 \\
0 & 1
\end{array}\right],\left[\begin{array}{cc}
0 & 1 \\
-1 & 0
\end{array}\right],\left[\begin{array}{cc}
-1 & 0 \\
0 & -1
\end{array}\right],\left[\begin{array}{cc}
0 & -1 \\
1 & 0
\end{array}\right],\left[\begin{array}{cc}
-1 & 0 \\
0 & 1
\end{array}\right],\left[\begin{array}{cc}
0 & -1 \\
-1 & 0
\end{array}\right],\left[\begin{array}{cc}
1 & 0 \\
0 & -1
\end{array}\right],\left[\begin{array}{ll}
0 & 1 \\
1 & 0
\end{array}\right]    \}$$

\begin{figure}
  \centering
  \includegraphics[width=.9\textwidth]{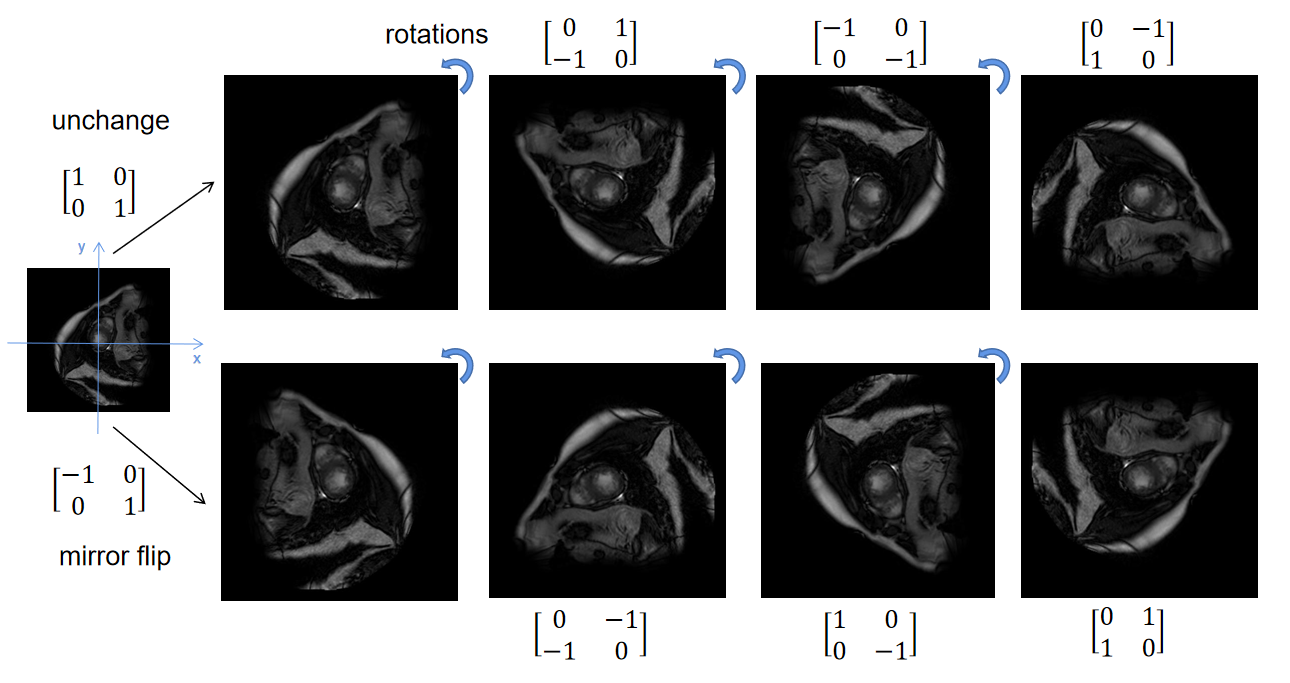} 
  \caption{8 possible orientations of the 2D image and their corresponding elements in $T$ } 
  \label{img1} 
\end{figure}

Actually, we can prove that the set $T$ is a subgroup of the group of second-order orthogonal matrices in which any two elements multiplied together still fall in the set, and all elements have inverse elements in the set. Then the orientation of the 2D image may have the following $8$ cases with equal probability (orientation only, see as Figure\ref{img1}). All $8$ of these cases can be obtained and recovered using elements in the set $T$.

\begin{figure}[H]
  \centering
  \includegraphics[width=.7\textwidth]{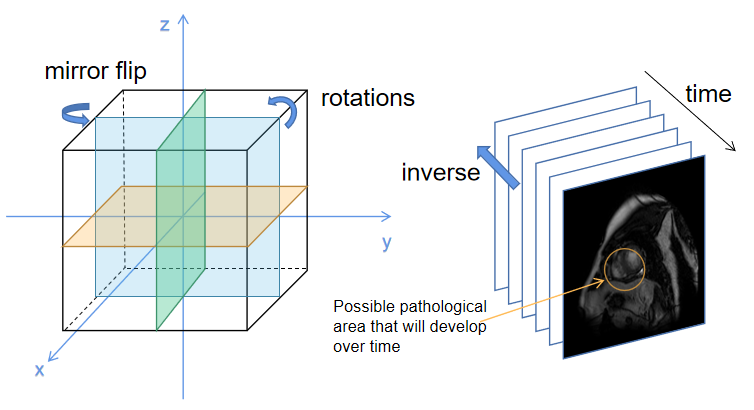} 
  \caption{Diagram of a 3D geometric transformation and a sample of serial images with added time dimension} 
  \label{img2} 
\end{figure}

We can easily extend this theory to 3D, where the mirror flip of an image needs to be along the xOy plane, the xOz plane, the yOz plane (Figure \ref{img2}). For a point $p_i(x_i,y_i,z_i)$ in the 3D rectangular coordinate system, we similarly use an $n\times3$ matrix $T_0$ to denote a point set $S$, the geometric transformations can also be represented by some $3\times3$ matrices. It can be shown that the set $T$ containing these $3\times3$ matrices has exactly the same properties as in 2D, except for the increase in the number of elements due to the increase in the number of dimensions. For 3D images, the orientation we focused on has $18$ cases. As in Figure \ref{img2}, we first let a 3D image flip along the xOy, xOz, yOz plane respectively, and then through the rotation, we get six orientations respectively, considering that the cube has six faces. 

This shows that a classification task can be designed for a 2D or 3D image to predict the specific geometric transformations that have acted on it and recover it to the positive direction of human vision by its inverse transformation.

\subsection{2D to 3D: Spatial Dimension or Time Dimension?}

In the previous subsection, we extended the geometric transformation from 2D to 3D based on the premise that the added dimension is still spatial. However, if we only wanted to explore the representation of geometric transformations by neural networks, then the increase in dimensionality would not be necessary; in other words, if geometric transformations can be recognized in 2D, then surely they can also be recognized in 3D.  

The situation would be different if the third dimension were the time dimension because the time and spatial dimensions do not equate. To be more precise, consider a video of a car slowly approaching and discretizing this video with continuous time into a number of frames; we need to segment the car in each frame. In medicine, consider a video that continuously records the development and spread of a lesion; sometimes, it would be a series of successive recordings of medical images every few months. Again, we need to segment the location where the lesion or necrosis is located, and we also need to focus on its evolution because of the appearance of the time dimension. In the above context, it is worthwhile to study the identification of geometric transformations in 3D (2 spatial dimensions and 1 time dimension) images, which we call serial images. 

Since the time has been discretized, it is, of course, possible to take all the images at different time points as input regardless of the time order. However, we have the idea that if the serial images are considered as one sample and input together, then remembering the region segmented at the previous time point will facilitate the segmentation at subsequent time points since the region is already determined, and only the extent of the spread needs to be predicted.  

Based on this idea, it makes sense to explore the recognition of 3D geometric transformations by neural networks if we want to augment the data by geometric transformations, such as image flips and rotations, clockwise and counterclockwise (Figure ef {img2}). However, it is essential to note that since the spatial and time dimensions are not equivalent, if we want to identify geometric transformations in serial images, only 16 cases will occur, including $8$ 2D image transformations in both clockwise and counterclockwise situations.

\section{Experiment}
\subsection{Data Generation and Pre-processing}

We train our model for a single modality on the MyoPS dataset \cite{data,dataa}, which provides the three-sequence CMR (LGE, T2, and bSSFP) images from 45 patients. Considering the dataset is relatively small, we divide all slices into two sub-sets, i.e., the training set and the test set, at the ratio of 80\% and 20\%.

\textbf{Geometric Transformations and Labels}  

For a single modality, we turn each slice of each 3D CMR image from 45 patients into a grayscale image and end up with \(174\) 2D images. As discussed in section 2.2, we then used \(8\) different geometric transformations for each image to obtain \(1392\) images. For an image \(X\), we get \(8\) samples:\( (G_y(X), G_y(\cdot)),y=1,…,8\), where \( G_y(\cdot) \) denotes the \(8\) different geometric transformations corresponding to those in \(T_0\), and also the label of the transformed image \(G_y(X)\). We use the ConvNet model \(F(\cdot)\) to predict the possible geometric transformation that may happen to the image.

\textbf{Data Augmentation}

\begin{figure}
  \centering
  \includegraphics[width=.72\textwidth]{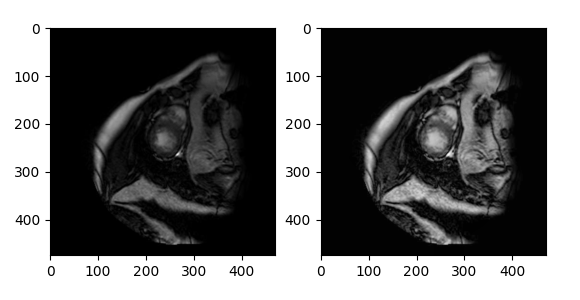} 
  \caption{A CMR image (bSSFP) before and after AHE augmentation} 
  \label{img3} 
\end{figure}

Before training, we used the Adaptive histogram equalization (AHE) method to process the above \(1392\) images, which is an image pre-processing technique used to improve contrast in images and was improved from traditional histogram equalization by Pizer \cite{AHE}. It computes several histograms, each corresponding to a distinct section of the image, and uses them to redistribute the image's luminance values. Figure \ref{img3} shows a clear comparison of an image before and after AHE augmentation.

Because we need to identify mirror flips and rotations of the image, we cannot use these transformations to augment the data. Instead, we used random cropping and scaling to select 70\% of the original training image and resize it all to $256\times256$ for the network. Although we cannot use a rotation of \(90\) degrees to enhance the images, we can still choose a much smaller angle, such as \(10\) degrees, to increase the number of images. This slight rotation does not change the label we are predicting but instead simulates the slight skewing of the images that may occur in practice. 

 \begin{figure}
  \centering
  \includegraphics[width=.95\textwidth]{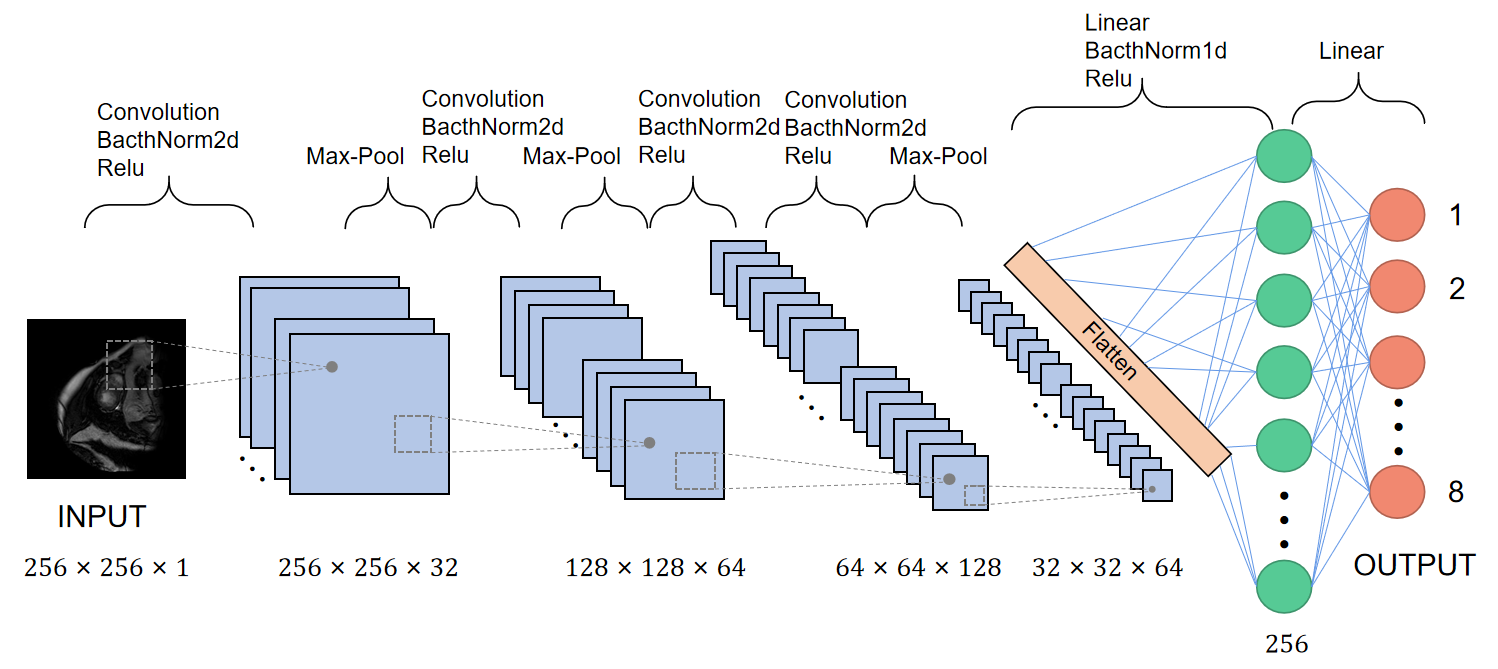} 
  \caption{The architecture of GeoNet and our inference process} 
  \label{img4} 
\end{figure}

\subsection{GeoNet}

According to Mumuni et al. \cite{CNN}, ConvNet is distinguished by its lightning-quick learning and inference speeds. Given these characteristics, it shows promise for future research and can be used as a general tool for addressing geometric-transformation invariant recognition issues. We will hereafter call a ConvNet model trained on the self-supervised task of geometric transformation recognition GeoNet model, whose architecture is designed based on AlexNet \cite{Alexnet} and RotNet \cite{Rotation}. 

We also implement the GeoNet models with Network-In-Network (NIN) architectures (Lin et al.,2013\cite{nin}), which is depicted in Figure\ref{img4} and comprises $6$ weight layers in total: it consists of $2$ convolutional layers each followed by a BatchNorm2d and a ReLU activation function and a max-pooling operation and followed by $2$ consecutive convolutional layers and a max-pooling layer, end up with a fully-connected layer with a final 8-class classifier. To solve the geometric transformation prediction task,  we train the network for $32$ epochs on the previously described augmented MyoPS with images of size $256\times256$, using an SGD optimizer with categorical cross entropy, a learning rate of $0.01$ and a batch size of $16$.

\subsection{Results}

\begin{figure}[htbp]
\centering  
\subfigure[Training and testing accuracy (T2)]{   
\begin{minipage}{8cm}
\centering    
\includegraphics[scale=0.6]{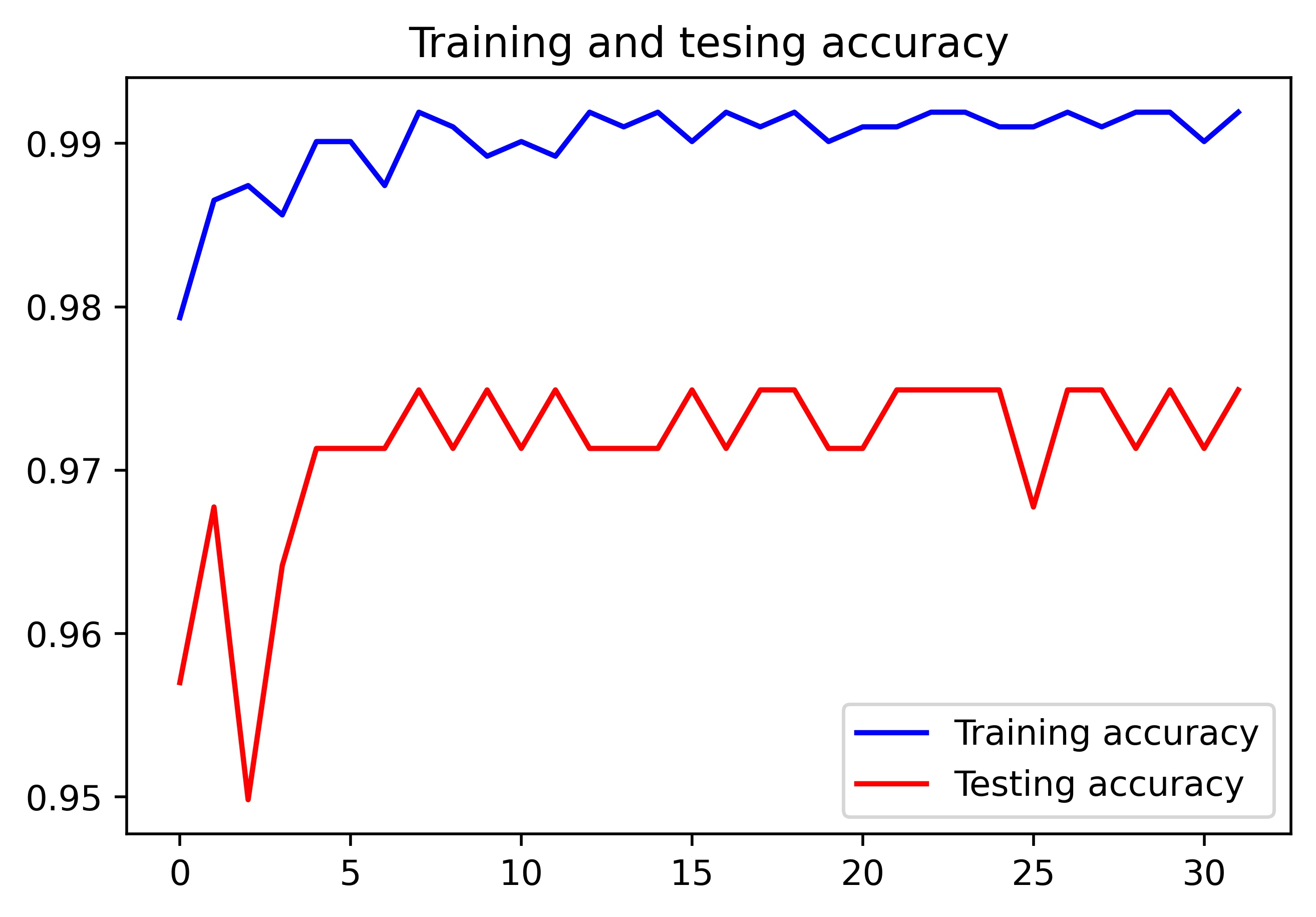}  
\label{img6.a} 
\end{minipage}
}
\subfigure[Training and testing loss (T2)]{ 
\begin{minipage}{8cm}
\centering    
\includegraphics[scale=0.6]{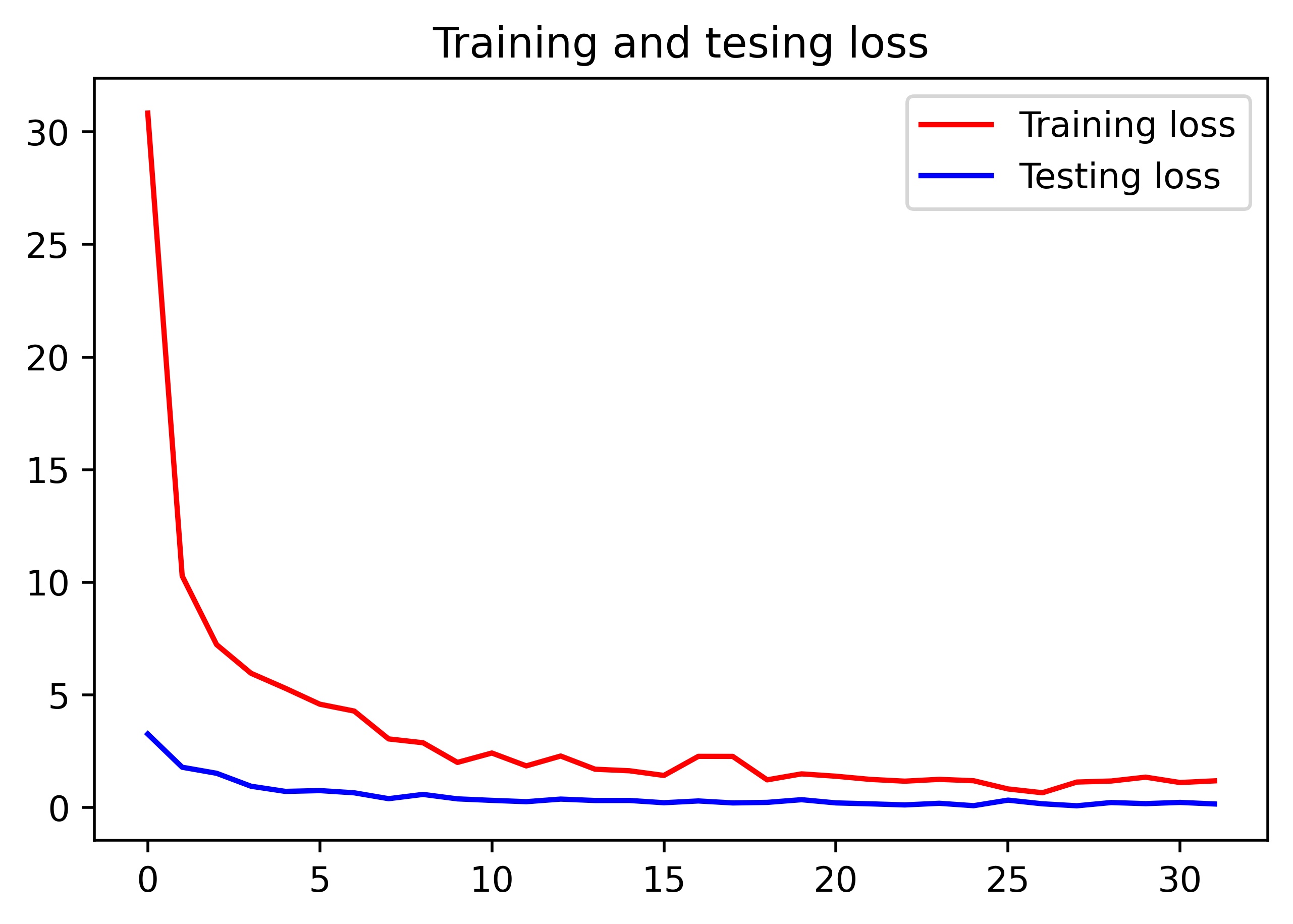}
\label{img6.b} 
\end{minipage}
}
\caption{Accuracy and Loss of a single modality}    
\label{img6}    
\end{figure}

We evaluated the performance of GeoNet while under training, using the metrics of accuracy and loss value obtained for the number of epochs, as you can see in Figure\ref{img6.a}, Figure\ref{img6.b}, the accuracy of the training and testing datasets on the model showed a progressive rise in value to an impressive range above 95\%. The loss values dropped significantly for both the training and testing curve to signify the discrepancies between images after different geometric transformations. Table \ref{tab1} shows the accuracy of different modalities.

\begin{table}[htbp]
  \centering
  \caption{Accuracy on the testing set of three modalities}
    \begin{tabular}{l|rrr}
    \toprule
          & \multicolumn{1}{l}{\textbf{bSSFP}} & \multicolumn{1}{l}{\textbf{LGE}} & \multicolumn{1}{l}{\textbf{T2}} \\
    \midrule
    \textbf{Accuracy} & 0.964 & 0.975 & 0.964 \\
    \bottomrule
    \end{tabular}%
  \label{tab1}%
\end{table}%

\begin{figure}
  \centering
  \includegraphics[width=1\textwidth]{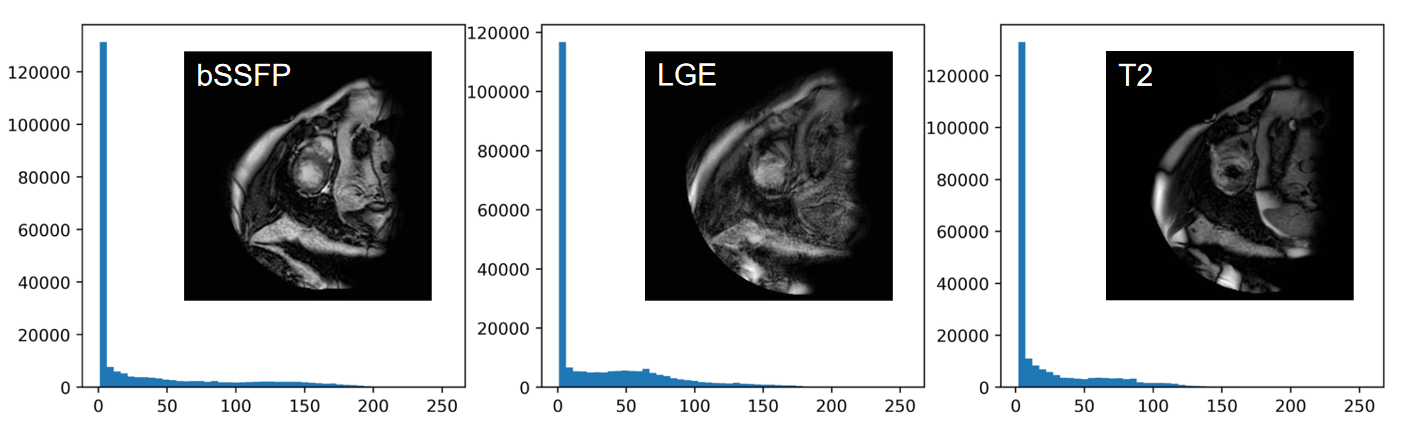} 
  \caption{Histograms of different modalities of the same slice of the same patient after AHE augmentation} 
  \label{img5} 
\end{figure}

We found that training with images of different modalities ended up with similar accuracy, which naturally raises the question of whether modality does not have a decisive influence on the recognition task of geometric transformations in images, considering that images of different modalities are similar after AHE augmentation in terms of essential features that play a decisive role, such as contour and shape. We can further corroborate from Figure \ref{img5}, the grey-scale histograms of different modalities of the same slice of the same patient after AHE augmentation are similar in value and distribution. 

\section{Conclusion}

We have described a set of 2D geometric transformations containing mirror flips and rotations in mathematical terms and extended it to 3D. We also highlighted the importance of adding a temporal dimension over a spatial one. We propose a simple self-supervised task that can easily predict the geometric transformation on CMR images. The experiment demonstrates that the geometric transformation recognition network GeoNet can recognize the orientation classification from multi-sequence CMR images. Our future research aims to explore the following points:

\begin{enumerate}
    \item We will combine the tasks of recognizing the geometric transformations of CMR images of different modalities.
    \item Using the feature maps, we will further compare and analyze the process of feature learning of the GeoNet and the network used in the supervised learning task.
    \item We will study the geometric transformation recognition and the segmentation tasks of serial medical images(3D medical images with 2 spatial dimensions and 1 time dimension).

\end{enumerate}

\bibliographystyle{unsrt}  


\end{document}